\documentclass[twocolumn,showpacs,preprintnumbers,amsmath,amssymb,prc]{revtex4}
\usepackage{graphicx}
\usepackage{dcolumn}
\usepackage{bm}
\begin{document}
\title{Further studies of multiplicity derivative in models of heavy ion collision at intermediate energies as a probe for phase transition}
\author{S Das Gupta$^{1}$, S. Mallik$^{2}$ and G. Chaudhuri$^{2}$}
\affiliation{$^1$Physics Department, McGill University, Montr{\'e}al, Canada H3A 2T8}
\affiliation{$^2$Physics Group, Variable Energy Cyclotron Centre, 1/AF Bidhan Nagar, Kolkata 700064, India}
\begin{abstract}
In conjunction with models, the experimental observable total multiplicity can be used to check if the data
contain the signature of phase transition and if it is first order.  Two of the models reach similar conclusions. The third one is quite different.
\end{abstract}
\maketitle

\section{Introduction}
This paper deals with identifying the order of phase transition from experimental data in intermediate energy
heavy ion collisions. We focus here on the total multiplicity $M$ resulting from central collisions of two heavy ions; $M$ is a function of the beam energy.  The derivative of $M$ with energy as a function of energy may go through a maximum.  In a previous paper \cite {Mallik16} we claimed that the appearance of this maximum is a signature of first order phase transition in the collision.  Absence of a maximum would imply there is no first order phase transition.  We used a canonical thermodynamic model (CTM) \cite {Das} to reach this conclusion.  As is usual in canonical model calculations the $M$ derivative is easiest to obtain with respect to temperature which can then be mapped in terms of energy.\\
\indent
The model is based on the ansatz that in heavy ion collisions a heated conglomeration of nucleons in an expanded
volume is formed.  Nucleons get grouped into various composites and the total number of composites plus monomers
is the total multiplicity $M$.  This system of particles can go through a phase transition. \cite{Siemens,Dasgupta_Phase_transition,Borderie2,Gross_phase_transition,Chomaz} The system is
characterised by a temperature $T$ and has an average energy $E$. At phase transition temperature, $C_v$, the derivative of energy
with respect to temperature $T$, goes through a maximum.  The quintessential problem is: how to recognise this
maximum experimentally.  Using CTM we found that the maximum of $\frac{dE}{dT}$ and the maximum of $\frac{dM}{dT}$ coincide. Since $\frac{dM}{dE}$ is experimentally accessible, signal for first order transition can be recognised.\\
\indent
Although the calculation of \cite{Mallik16} was done with CTM only, we expect similar results with microcanonical
models \cite{Bondorf1,Gross}. The basic physics assumptions are the same.  In examples where microcanonical and canonical calculations were compared \cite{Chaudhuri_plb} they were found to be very close.  We note in passing that both canonical and microcanonical models are found to give in general very good fits to experimental data.\\
\indent
Here we will examine features of $M$ derivatives for models different from standard thermodynamic models. Of
particular interest \cite{Referee} is the percolation model \cite{Campi,Bauer} which was extensively used in the
past to establish a link between experimental data and phase transition.  In the context of the present work
percolation results will be very interesting since percolation is a model of continuous phase transition.  We
will next examine the M derivative in the lattice gas model which uses similar geometry as percolation
but is much more elaborate with the insertion of a Hamiltonian.  First order transition is possible
here \cite{Pan,Samaddar,Chomaz}.\\
\\
\section{Total Multiplicity and its derivative in Percolation Model}
\begin{figure}[b]
\includegraphics[width=5.5cm,keepaspectratio=true]{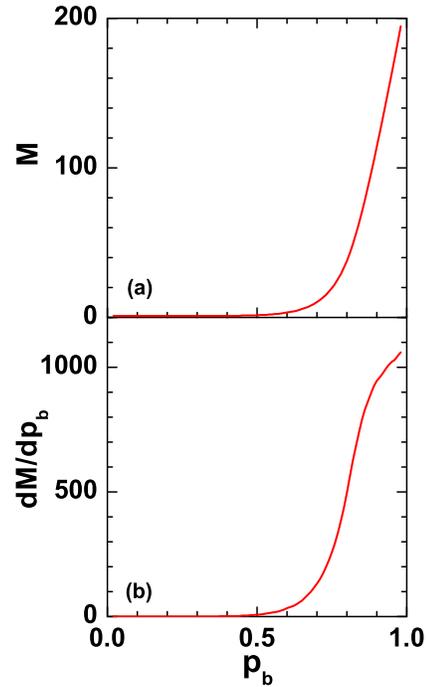}
\caption{(Color Online) Variation of $M$ (upper panel) and $\frac{dM}{dp_b}$ (lower panel) with $p_b$ obtained from bond percolation model for a system of $6^3$ nucleons.}
\end{figure}
\begin{figure}[b]
\includegraphics[width=5.5cm,keepaspectratio=true]{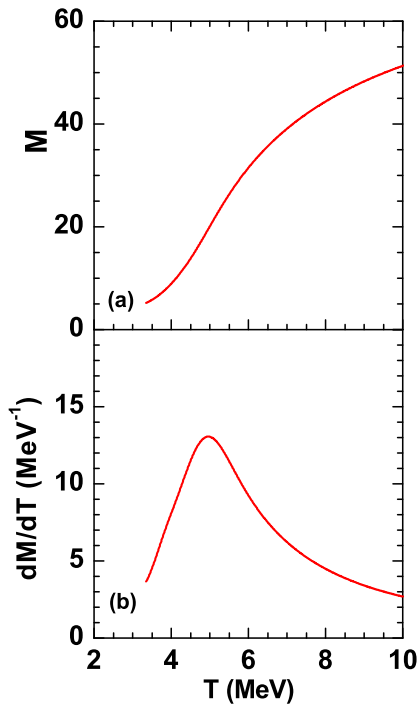}
\caption{(Color Online) Variation of $M$ (upper panel) and $\frac{dM}{dT}$ (lower panel) with $T$ obtained from CTM for fragmenting system having $Z$=82 and $N$=126.}
\end{figure}
\begin{figure}[t]
\includegraphics[width=5.5cm,keepaspectratio=true]{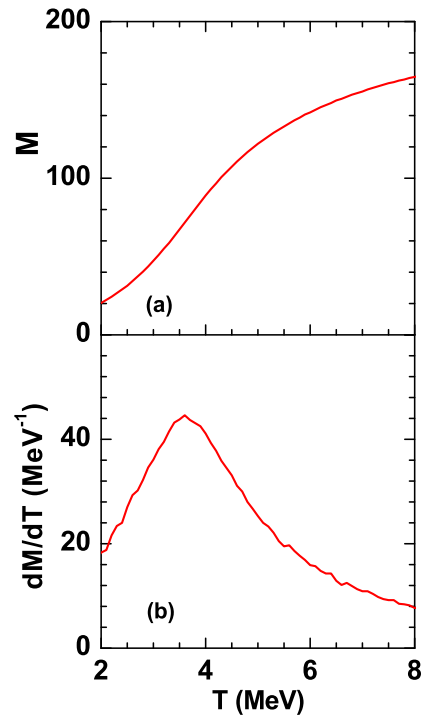}
\caption{(Color Online) Variation of $M$ (upper panel) and $\frac{dM}{dT}$ (lower panel) with $T$ obtained from lattice gas model for fragmenting system having $Z$=82 and $N$=126}
\end{figure}
We consider a system of $6^3$ nucleons in bond percolation model.  The model does not distinguish between neutrons
and protons.  There are $6^3$ boxes and each box contains one nucleon.  Nearest neighbours (these have a common wall) can bind together with a probability $p_s$.  If $p_s$ is 1 there is just one nucleus with $6^3$ nucleons and $M$=1. If $p_s$ is 0 there are $6^3$ monomers and $M$=$6^3$  For intermediate values of $p_s$ nucleons can group into several composites.  For an "event" this is obtained by Monte-Carlo sampling.  Let the average number of clusters of $a$ nucleons be $n_a$.  Then $M$=$\sum_a n_a$.  In bond percolation model there is just one parameter $p_s$.  Thus we can plot $M$ against $p_s$ and examine the $M$ derivative.  Instead of plotting $M$ against $p_s$ we plot $M$ against $p_b$=$1-p_s$ which is the bond breaking probability.  If $p_b$ is $0$, then $6^3$ nucleons appear as one cluster and $M$=1.  If $p_b$ is 1 then we have $6^3$ monomers and $M$=$6^3$.
\indent
Fig.1 plots $M$ and $\frac{dM}{dp_b}$ in the range of $p_b$ 0 to 1.  For reference in Fig.2 we have plotted $M$ and $\frac{dM}{dT}$ as was obtained in CTM \cite{Mallik16}.  Both $M$ and $M$ derivatives are very different in the two models.  Percolation model has no first order phase transition and as conjectured before \cite{Mallik16} there is no maximum in the $M$ derivative.  Also note that the CTM calculations are quite realistic.  The inputs were liquid drop model energies for composites.  Coulomb interaction between composite are included approximately. If one omits Coulomb interaction between different composites the maximum in $M$ derivative becomes sharper.\\
\indent
The well-known function of $p_b$ that is normally used is not $M(p_b)$ but a second moment function $m_2(p_b)$.  That function has a maximum at about $p_b$=0.8 (equivalently $p_s$=0.2).  We will use that function in section IV.\\
\section{Total Multiplicity and its derivative in the Lattice Gas Model}
Next topic we deal with is $M$ derivative in the Lattice Gas Model. This is shown in Fig.3.  Here $M$ and its
derivative are plotted against the temperature $T$. The Lattice Gas Model is considerably more complicated than the percolation model but expositions of the model exist \cite{Pan,Samaddar,Chomaz} and we refer to \cite{Samaddar} for details.  Let $A=N+Z$ be the number of nucleons in the system that dissociates.  We consider $D^3$ cubic boxes where each cubic box has volume $(1.0/0.16)fm^3$. $D^3$ is larger than $A$ (they have the same value in bond
percolation model). Here $D^3/A=V_f/V_0$ where $V_0$ is the normal volume of a nucleus with $A$ nucleons and $V_f$
is the freeze-out volume where partitioning of nucleons into clusters is computed.  For nuclear forces one adopts
nearest neighbor interactions.  Following normal practice, we use neutron-proton interactions $v_{np}$=-5.33 MeV and set $v_{nn}=v_{pp}$=0.0. Coulomb interaction between protons is included.  Each cube can contain 1 or 0 nucleon. There is a very large number of configurations that are possible (a configuration designates which cubes are occupied by neutrons, which by protons and which are empty; we sometimes call a configuration an event).  Each configuration has an energy.  If a temperature is specified, the occupation probability of each configuration is proportional to its energy: P$\propto$exp(-E/T).  This is achieved by Monte-Carlo sampling using Metropolis
algorithm.\\
\begin{figure}[t]
\includegraphics[width=6.5cm,keepaspectratio=true]{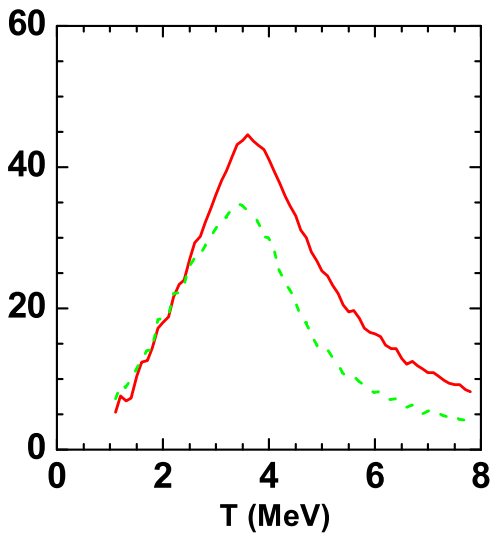}
\caption{(Color online) Variation of d$M$/d$T$ (red solid lines) and $C_v$ (green dashed lines) with temperature from lattice gas model at $D$=8 (see text) for fragmenting system having $Z$=82 and $N$=126. To draw d$M$/d$T$ and $C_v$ in the same scale, $C_v$ is normalised by a factor of 1$/$10; d$M$/d$T$ is unit of MeV$^{-1}$.}
\end{figure}
\indent
Calculation of clusters need further work.  Once an event is chosen we ascribe to each nucleon a momentum.
Momentum of each nucleon is picked by Monte-Carlo sampling of a Maxwell-Boltzmann distribution for the prescribed
temperature T.  Two neighboring nucleons are part of the same cluster if $\vec{P}_r^2/2\mu+\epsilon<0$
where $\epsilon$
is $v_{np}$ or $v_{nn}$ or $v_{pp}$.  Here $\vec{P}_r$ is the relative momentum of the two nucleons and $\mu$ is
the reduced mass.  If nucleon $i$ is bound with nucleon $j$ and $j$ with $k$ then $i, j, k$ are part of the same
cluster.  At each temperature we calculate 50,000 events to obtain average energy
$<E>$ and average multiplicity $n_a$
(where $a$ is the mass number of the cluster) of all clusters.  A cluster with 1 nucleon is a monomer, one with 2
nucleons is a dimer and so on.  The total multiplicity is $M=\sum n_a$
and $\sum an_a=A$ where $A=N+Z$ is the mass number of the dissociating system.
Plots of $dM/dT$ and $d<E>/dT$ are shown in Fig 4.  Note that $c_v$ goes through a maximum at some temperature
which is a hallmark of first order phase transition and this occurs at the same temperature where $dM/dT$
maximises.  This is remarkably different from percolation model results but very similar to CTM results of
\cite{Mallik16} corroborating the evidence that the appearance of a maximum in $dM/dT$ is indicative of a first
order phase transition.\\
\begin{figure}[b]
\includegraphics[width=7.0cm,keepaspectratio=true]{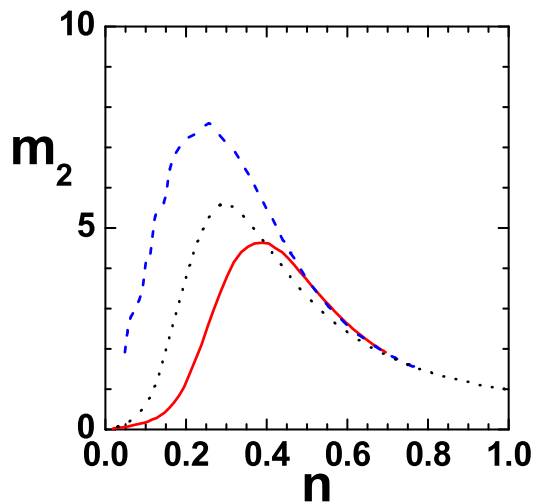}
\caption{(Color online) Variation of $m_2$ with $n$ calculated from from lattice gas model at $D$=7 (red solid line) and $D$=8 (blue dashed line) and percolation model (black dotted line) for fragmenting system having $Z$=82 and $N$=126.}
\end{figure}
\section{The second moment $m_2$ in the models}
Although the percolation model curves that we have shown above are even qualitatively different from those
emerging from the lattice gas model and the CTM, there is one curve that is similar and was used a great
deal when percolation was the only available microscopic model to link experimental multifragmentation data
to phase transition.  We will call this the second moment curve $m_2$.  Consider the percolation curve of Fig.1 where we chose
the dissociating system to consist of $A$=216 nucleons.  Define reduced multiplicity $n=M/A$
where $M$ is the total multiplicity and $A$ is the mass of the dissociating system;  $n$ varies from
$1/A\approx 0$ to 1 as $p_b$ goes from 0 to 1.  We expect $M$  to increase if more energy is pumped in
the system.  For example in counter experiments one can gate on central collisions and vary the beam energy.
In emulsion experiments \cite{Waddington,Campi} there is no selection on the impact parameter and in collisions
at different impact parameters different amounts of energies are pumped in for multifragmentation. For our illustration purposes we consider central collisions for two models, the percolation and the lattice gas model in a range of energies. For these we will plot $m_2$ as a function of $n$.  Define $m_2$ by
\begin{equation}
m_2=\frac{\bigg{[}\sum a^2n_a-a_{max}^2\bigg{]}}{A}
\end{equation}
We denote by $a_{max}$ the largest cluster in an event. For percolation we pick a $p_b$ and get $n$, $n_a$ and $a^2_{max}$ by averaging over 50,000 events.  Thus we can plot $m_2$ against $n$. For lattice gas model we take 50,000 events at each temperature and follow the same procedure. The $m_2$ curves are given in Fig. 5. Note that $m_2$ curves for the lattice gas model and the percolation models are quite similar and from experimental data (which can be fitted only approximately) one could choose either a percolation model or a lattice gas model. But the models have in fact even different orders of transition. If we defined $m_1=\sum an_a/A$ the answer is identical in both the models with value 1; just a straight line  with a value 1 for all n's. One can build a little bit of structure if we defined $m_1=[\sum an_a-a_{max}]/A$ but the $m_2$ is the first interesting quantity, though not a confirmatory signal.\\
\begin{figure}
\includegraphics[width=5.5cm,keepaspectratio=true]{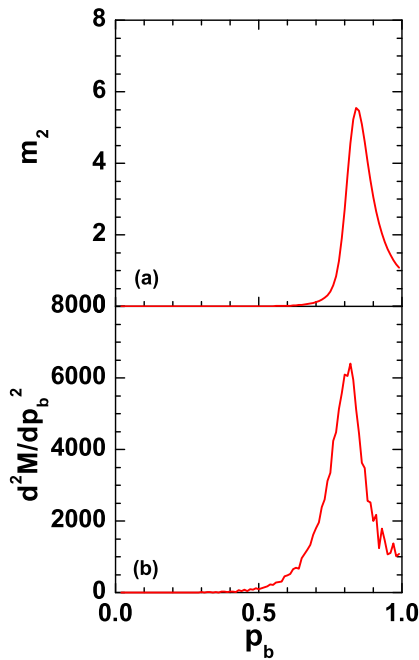}
\caption{(Color Online) Variation of $m_2$ (upper panel) and $\frac{d^2M}{dp_b^2}$ (lower panel) with $p_b$ obtained from bond percolation model for a system of $6^3$ nucleons.}
\end{figure}
\section{Back to Percolation}
In the previous section we compared $m_2$ obtained from the lattice gas model and the percolation model. For that purpose it was convenient to plot $m_2$ as a function of $n$. Now we concentrate on percolation model only and it will be more convenient to draw $m_2$ as a function of $p_b$. In Fig.5 we drew a curve of $m_2$ as a function of $n$.  In this section it will be more convenient to draw $m_2$ as a function of $p_b$.  In Fig.1 of section II we drew a curve of both $M$ and $dM/dp_b$.  We now draw a curve of $d^2M/dp_b^2$ and compare with $m_2(p_b)$ in Fig.6.  The similarity of the two is remarkable. The mathematics in computing $m_2$ and the second derivative are very different.  One is tempted to conclude that the second derivative of $M$ having a maximum is an indication that this is a case of second order phase transition.\\
\section{Discussion}
Recognition of phase transition in intermediate energy collisions has been an interesting and intriguing problem
of long standing.  A popular approach was to try to best fit individual multiplicity $n_a$ to a form suggestive
of critical phenomenon: $n_a=a^{-\tau}f(a^{\sigma}((T-T_c))$ \cite{Bauer,Campi,Elliot1,Elliot2,Scharenberg}.
It is impossible to get a very good fit as the masses $a$ need to be big for the model to work and in heavy ion
collisions in the lab fragment sizes are limited.  As fits are only approximate very different models can give similar quality fits.  Thus the conclusions are ambiguous.  Here we have specialised to an observable which is very feasible to scan and will give an unambiguous answer.\\
\indent
In addition we have identified an interesting feature of $M$ in percolation model which was not recognised before.\\

\end{document}